# Title: Unlock giant nonreciprocity via multi-valued behavior of non-Hermitian zero-index materials

**Authors:** Yang Li[1,2]*†, Yueyang Liu[2]†, Yucong Yang[3,4]†, Tianyi Zhang[2]†, Jianfeng Chen[5], Tian Dong[2], Fulong Shi[5], Phatham Loahavilai[2], Tianchi Zhang[3,4], Di Wu[3,4], Zixuan Wei[3,4], Dengfu Deng[3,4], Jun Qin[3,4], Longjiang Deng[3,4]*, Cheng-Wei Qiu[5]*, Lei Bi[3,4]*

**Affiliation**

Yang Li[1,2]*, Yueyang Liu[2], Tianyi Zhang[2], Tian Dong[2], Phatham Loahavilai[2]

[1]State Key Laboratory of Optoelectronic Materials and Technologies, School of Electronics and Information Technology, Sun Yat-sen University, Guangzhou 510275, China.

[2]State Key Laboratory of Precision Measurement Technology and Instruments, Department of Precision Instrument, Tsinghua University, Beijing, 100084 China.

*Corresponding author. Email: liyang328@mail.sysu.edu.cn

Yucong Yang[3,4], Tianchi Zhang[3,4], Di Wu[3,4], Zixuan Wei[3,4], Dengfu Deng[3,4], Jun Qin[3,4], Longjiang Deng[3,4]*, Lei Bi[3,4]*

[3]National Engineering Research Center of Electromagnetic Radiation Control Materials, Key Laboratory of Multi-spectral Absorbing Materials and Structures of Ministry of Education, University of Electronic Science and Technology of China, Chengdu, 610054 P. R. China

[4]Key Laboratory of Multi-spectral Absorbing Materials and Structures of Ministry of Education, University of Electronic Science and Technology of China, Chengdu 610054, China.

*Corresponding author. Email: denglj@uestc.edu.cn; bilei@uestc.edu.cn

Jianfeng Chen[5], Fulong Shi[5], Cheng-Wei Qiu[5]*

[5]Department of Electrical and Computer Engineering, National University of Singapore, Singapore, 117583 Singapore.

*Corresponding author. Email: chengwei.qiu@nus.edu.sg

**Abstract**

Although Einstein's field equations are time-independent, the multivalued feature of the horizon of a blackhole naturally enables the one-way transmission, leading to the strong arrow of time from the time-independent gravitational interaction. Here we experimentally demonstrate a photonic analogue of this principle and reveal the infinite nonreciprocity of the time-reversal-symmetric Maxwell equations. By designing a non-Hermitian zero-index magneto-optical metawaveguide, we introduce multivalued feature to this metawaveguide's complex eigenspace via an exceptional point with non-zero residue, bringing nonlocal, path-dependent historical memory to the system. Hence, a weak magneto-optical response can direct forward and backward waves to two photonic branches with largely distinct momenta and losses, leading to the optical nonreciprocity far beyond the limitation imposed by the magneto-optical material. We fabricated an a-Si/Ce:YIG metawaveguide, achieving nonreciprocal phase shift of 47.78 rad/mm and nonreciprocal loss of 53.9 dB/mm near 1575 nm, exceeding state-of-the-art nonreciprocal devices by an order of magnitude. Our principle universally applies from microwave to visible frequencies, leading to compact isolators, circulators, and sensors. Our principle can also be extended to nonreciprocal acoustic, elastic, and thermal systems. The proposed new paradigm — geometry-based strong arrow of time in covariant and reversible physical systems — has broad implications in many disciplines including string theory, cosmology, and astronomy.

**Introduction**

Most equations describing basic interactions, such as the Einstein field equations and Maxwell's equations in covariant form[1,2], unify time and space under a four-dimensional framework, resulting in elegant and simple symmetry to these basic physical laws. Hence, the basic equations of gravitational interactions and electromagnetic interactions are both time-reversal symmetric.

However, black hole — as an extreme gravitational system — exhibits a strong arrow of time under time-reversal symmetric gravitational interactions (Fig. 1a). The black hole's horizon accepts the incoming particles but only allows the Hawking radiation along the opposite direction[3-5], featuring one-way information transmission. This peculiar physical behavior originates from the logarithmic branch point of the black hole's horizon — the non-zero residue introduces the global multi-valued feature and path memory while encircling the branch point. As shown in Fig. 1b, when we apply the time-reversal symmetric transform to the process that the wave function falls into the horizon, the wave function enters the new branches (red curves in Fig. 1b) instead of returning to the original Riemann surface. Such a multi-valued feature breaks the single-frequency condition, leading to the thermal emission — the Hawking radiation — under the theoretical framework of second quantization (see supplementary information, section S1, for detailed discussion). The Hawking radiation's temperature is determined by the encircling phase of the logarithmic branch point. This encircling phase is solely determined by the black hole's mass and hence excludes any information of the source wave function.

Similar to the Einstein field equations, Maxwell equations can also be expressed in covariant form. This intrinsic space-time symmetry leads to the dependence of electromagnetic nonreciprocity on the external modulations that break the time-reversal symmetry. These external modulations include magneto-optical modulation[6-11], spatial-temporal modulation[12-18], and nonlinear modulation[19-25]. Hence, electromagnetic nonreciprocal devices' nonreciprocity is limited by the materials' intrinsically weak optical nonreciprocal properties, including the off-diagonal term of the magneto-optical material's permittivity tensor, electro-optic/acoustic-optic coefficient, and the nonlinear coefficient. Consequently, the difference between forward and backward propagations is extremely small (Figs.1 c and e), leading to device lengths up to hundreds or thousands of wavelengths. These nonreciprocal devices are about one order larger than other semiconductor photonic devices, preventing high-density integration.

Here we introduce the arrow of time originating from the multivalued geometry of gravitational interaction to the electromagnetic interaction. By introducing ½-residue branch point to a non-Hermitian zero-index magneto-optical metawaveguide[26-34], we bring the multi-valued feature and path memory while encircling the branch point to the complex eigenspace of the metawaveguide. As shown in Figs. 1 d and f, under a weak magneto-optical modulation, the backward wave enters a branch strongly coupled to the radiation losses (radiation continuum) rather than traveling along the branch coupled to the guided modes (bound state) where the forward wave propagates. The radiation-continuum branch and bound-state branch have largely distinct momenta and losses (see supplementary information, section S2, for detailed discussion), leading to a giant optical nonreciprocal phase shift (NRPS) and a nonreciprocal loss (NRL). Consequently, this nonreciprocity is no longer limited by the materials' intrinsically weak nonreciprocal properties in the optical regime. Instead, this nonreciprocity is determined by the momentum/loss difference between the radiation-continuum branch and the bound-state branch, enabling nonreciprocity enhancement via device engineering. Hence, this nonreciprocity can be, in principle, infinite!

The nonreciprocal performance is typically evaluated by the NRPS and NRL, quantifying the phase and loss differences between forward and backward propagations. According to our theory, we designed and fabricated a silicon/cerium-substituted yttrium iron garnet (Si/Ce:YIG) non-Hermitian zero-index metawaveguide supporting a branch point with a ½-residue, depicting the record-high NRPS of 47.78 rad/mm at 1579 nm and NRL of 53.9 dB/mm at 1575 nm. These values are over one order of magnitude higher than those of state-of-the-art nonreciprocal photonic waveguides with different mechanisms operating at similar wavelengths (Fig. 1g), resulting in significantly shorter device footprints.

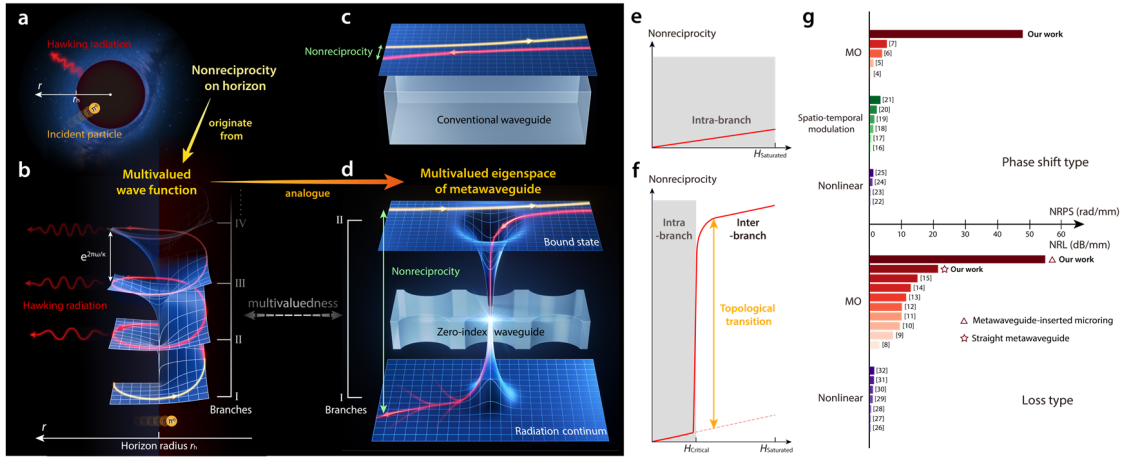

**Fig. 1. Comparison between a conventional nonreciprocal waveguide and a multi-value behavior-based nonreciprocal zero-index metawaveguide. a,** Schematic of a black hole. **b,** Schematic of the Riemann surface near the black hole horizon. **c,** Schematic of the forward and backward propagating waves in a conventional nonreciprocal waveguide. **d,** Schematic of the forward and backward propagating waves in a nonreciprocal zero-index metawaveguide. **e,** The nonreciprocity of a conventional waveguide as a function of the applied magnetic field. **f,** The nonreciprocity of a zero-index metawaveguide as a function of the applied magnetic field. **g,** Comparison of NRPS and NRL between zero-index metawaveguide and conventional nonreciprocal waveguides[7,9,13-16,19-22,25,38-48]. The devices based on the same mechanism are labeled in the same color palette. MO: magneto-optical.

**Theory and design**

We designed a metawaveguide whose impedance-matched zero index corresponds to a quasi-Dirac cone dispersion with a tiny bandgap, consisting of a lower-loss branch and a higher-loss branch at the Brillouin-zone center (Fig. 2c)[26, 35-37]. The design is a metawaveguide consisting of a one-dimensional array of silicon bowtie structures on a Ce:YIG film grown on a Ca, Mg, Zr substituted gadolinium gallium oxide (sGGG) substrate (Fig. 2a). Such a metawaveguide shows multivalued behavior via two photonic branches with largely distinct wavevectors and losses, enabling giant NRPS and NRL. To balance the trade-offs between total propagation loss, the wavevector and loss differences between the two branches, and fabrication yield, we chose the geometric parameters as follows: $a$=777 nm, $b$=450 nm, $h$=230 nm, $r_a$=460 nm, and $r_b$=190 nm (see Methods and supplementary information, sections S3 and S4, for detailed discussion). The simulated a-Si/Ce:YIG metawaveguide shows a propagation loss of ~8 dB/mm near the epsilon-mu-near-zero (EMNZ) wavelength (Fig. 2b). In this design, the radiation loss of the higher-loss branch is twice that of the lower-loss branch (Fig. 2d), causing the simulated radiative loss.

We numerically computed the real and imaginary parts of the eigenvalues of the metawaveguide. As shown in Fig. 2c, the real parts of the lower-loss branch and the higher-loss branch correspond to the monopole-mode and dipole-mode field distributions, respectively, over a unit cell. These two branches are almost degenerated near the Brillouin-zone center, featuring two flat bands separated by a tiny bandgap. The size of this bandgap determines the bandwidth of the NRPS and NRL, as well as the critical magnetic field required to achieve the topological transition (Fig. 1f, see supplementary information, section S2, for detailed discussion). Hence, there is a tradeoff between the bandwidth of the enhanced NRPS and the strength of the magnetic field required to achieve the topological transition. In our design, the lower-loss branch and the higher-loss branch are engineered to be nearly degenerate, allowing the topological transition to occur under a relatively weak magnetic field of 37 Oe (see supplementary information, sections S3 and S5, for detailed information). In this case, the imaginary parts of the eigenvalues form the upper and lower arcs of a ring for the higher-loss branch and the lower-loss branch, respectively, as shown in Fig. 2d. Such two branches with largely distinct losses cannot guide the forward and backward propagations with a large NRL because none of these two branches has a consistent group-velocity direction (Fig. 2c).

The topological transition enables forward and backward propagations along two branches with a momentum difference, leading to a giant NRPS. The bandstructure under an applied magnetic field exceeding the critical value is shown in Fig. 2e. Once the topological transition happens, the forward and backward branches across the exceptional point (EP) with a finite group velocity, thereby unlocking the large momentum associated with the EP. To quantify the momentum difference between the backward and forward branches, we must mirror the backward branch relative to the vertical zero-momentum line because the positive direction of momentum is defined by the group velocity. This momentum difference between forward and backward propagating waves leads to a prominent peak in the NRPS spectrum (Figs. 2e and 2f) (see supplementary information, section S2, for detailed discussion).

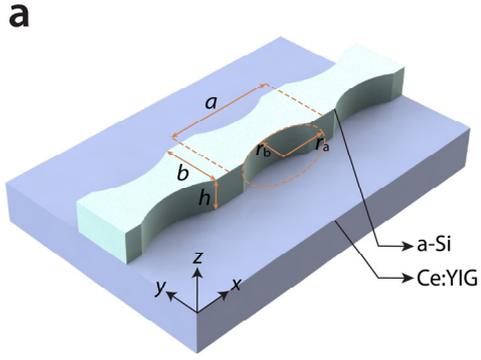
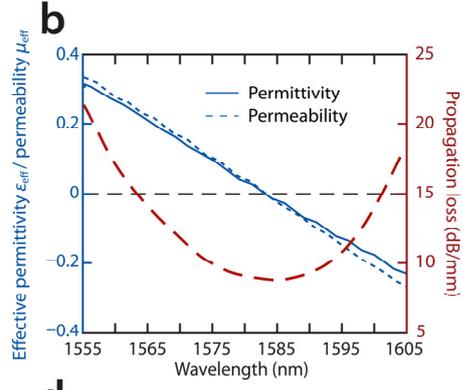
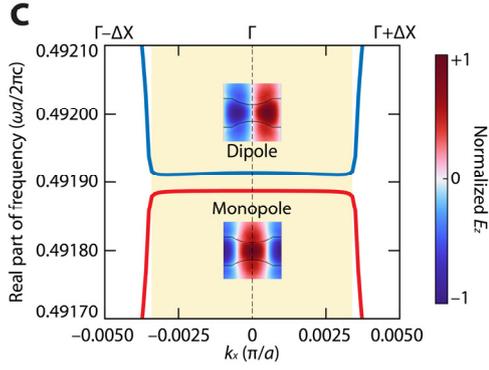
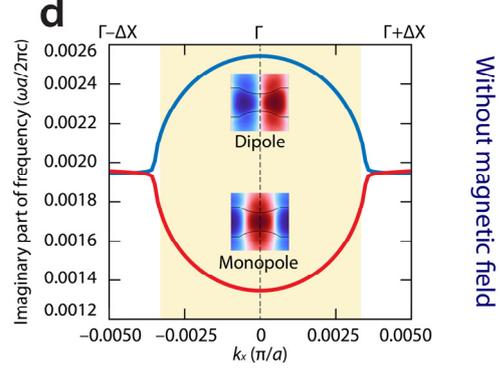
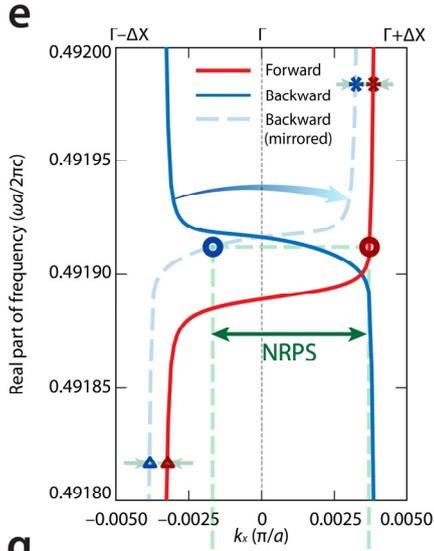
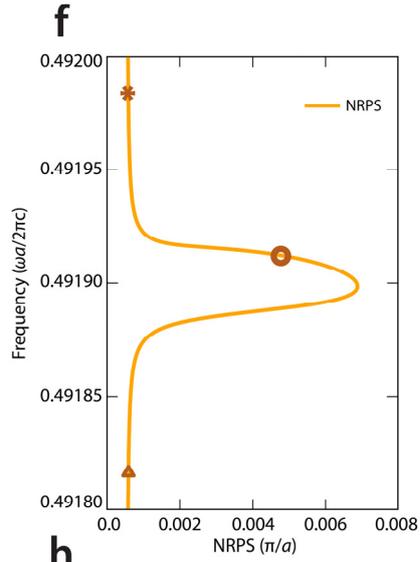
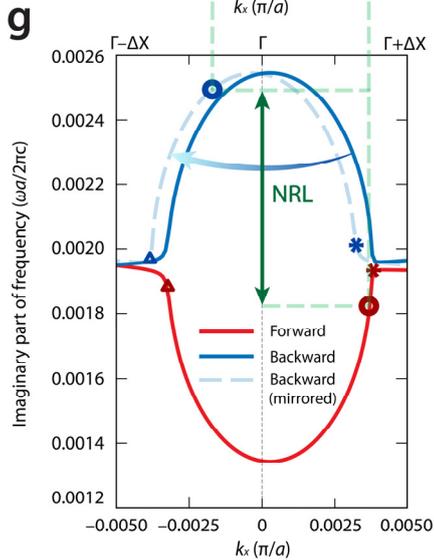
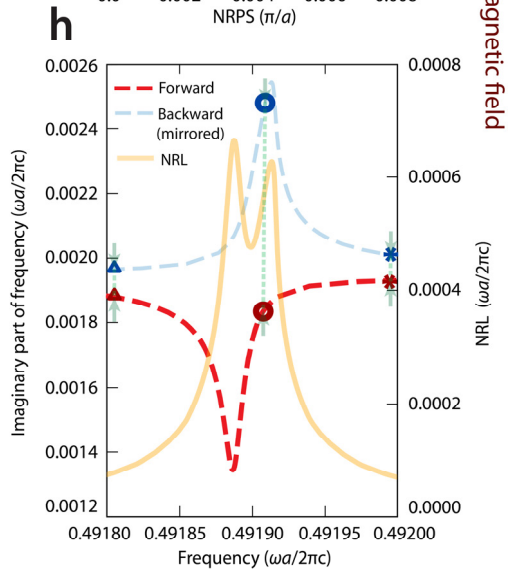

**Fig. 2. Schematic, effective constitutive parameters, propagation loss, bandstructures, NRPS spectrum, forward and backward loss spectra, and NRL spectrum of the metawaveguide. a.** Schematic of the metawaveguide. Geometric parameters: $a$=777 nm, $b$=450 nm, $h$=230 nm, $r_a$=460 nm, and $r_b$=190 nm. **b.** Simulated propagation loss and real parts of the effective permittivity and permeability of the metawaveguide. **c, d.** Real **(c)** and imaginary **(d)** parts of the bandstructure near the Γ point under zero magnetic field. **e, g.** Real **(e)** and imaginary **(g)** parts of the bandstructure near the Γ point under a magnetic field above the critical value. **f.** NRPS spectrum. **h.** Forward and backward propagation losses spectra and the resulting NRL spectrum.

The topological transition also directs backward and forward propagations to the higher-loss branch and the lower-loss branch, respectively, leading to a large NRL (Figs. 2g and 2h). To quantify this NRL, we map the imaginary parts of the forward and backward eigenvalues to their corresponding propagation-loss spectra via their associated wavenumbers (green lines between Fig. 2e and Fig. 2g). The difference in the imaginary parts of the frequencies in Fig. 2h reveals a pronounced asymmetry in the forward and backward propagation-loss spectra near the degeneracy-point frequency, corresponding to a giant NRL.

The topological transition-unlocked giant NRPS and NRL can be interpreted by the magnetically induced transition of the representative curves across the EP in the conformally mapped frequency space. In this space, the two representative curves lie on two Riemannian surfaces with distinct losses (Figs. 3a, b). These two representative curves' projection on the horizontal plane is a parabola whose vertex is located in the third quadrant (Fig. 3c). Hence, as the applied magnetic field increases, this parabola's opening broadens (see Movie S4 and supplementary information, section S2). When the applied magnetic field reaches its critical value, the parabola encircles the EP with a ½-residue, leading to the topological transition from the 0-residue trivial state to the

½-residue nontrivial state (Fig. 1f). During this transition, half of the representative curve on the lower-loss (higher-loss) Riemannian surface jumps from the lower (upper) Riemannian surface to the upper (lower) one (Figs. 3d and g, see Movie S1), unifying the group velocity's sign along each representative curve. Correspondingly, the right part of the photonic branch flips from the lower (upper) to the upper (lower) half of the photonic bandstructure (Figs. 3e and h), transforming a photonic bandgap (Fig. 3f) to the crossing of the forward and backward photonic branches at a topologically protected degeneracy point (Fig. i, see Movie S3).

Hence, in the complex space with non-zero residue-induced multi-valued feature, we introduced path memory encircling the branch point, leading to the forward and backward propagations along distinct branches. This is the electromagnetic counterpart of the geometric arrow of time on the black hole's horizon (see supplementary information, section S1).

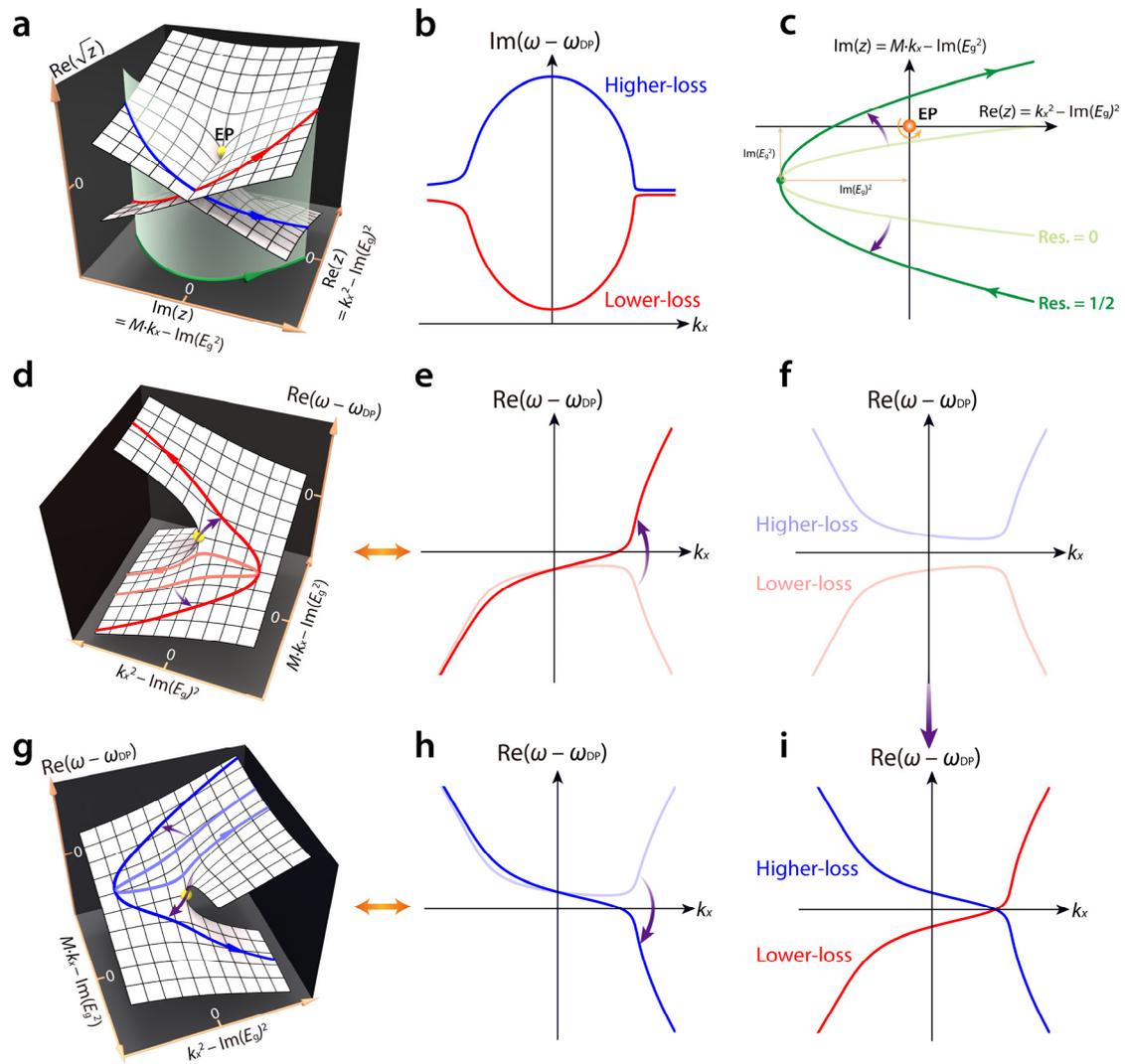

**Fig. 3. The magnetically induced topological transition of the zero-index metawaveguide. a.** Real representative curves on the Riemannian surface in the 3D conformally mapped frequency space. Red and blue representative curves with lower loss and higher loss, respectively. **b.** Imaginary parts of the bandstructure near the Γ point. **c.** The topological transition of representative points via the exceptional point (EP) in the 2D conformally mapped frequency space. **d, g.** The topological transitions of the representative curves via the EP. DP: Dirac point. **e, h.** The topological transition of each photonic branch. **f, i.** The topological transition of two photonic branches.

**Experimental results**

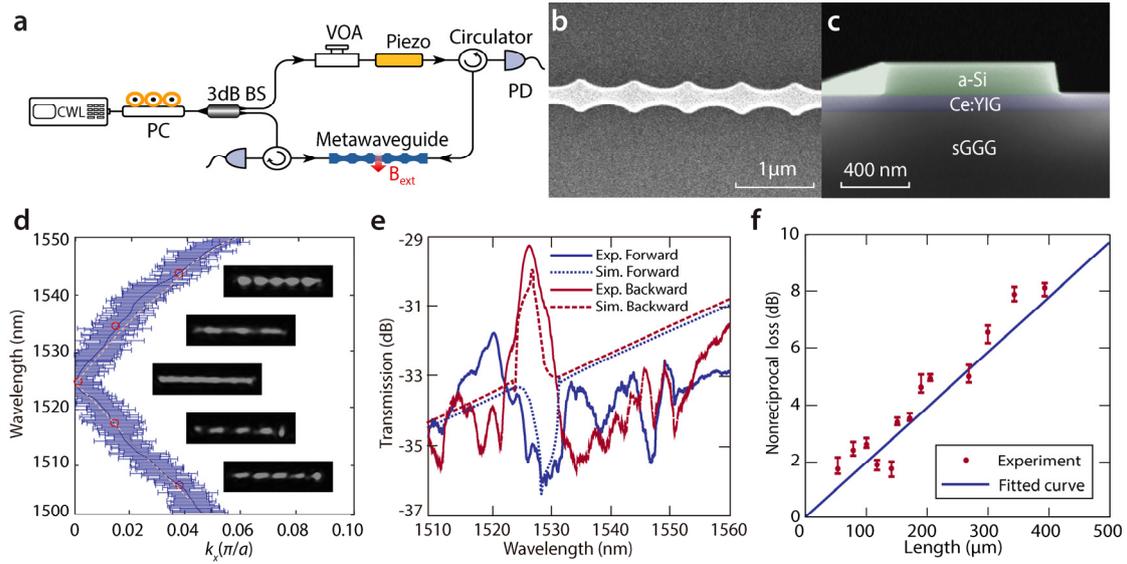

**Fig. 4. NRL measurement of an a-Si/Ce:YIG metawaveguide in the near infrared regime. a.** Schematic of the experimental setup for measuring the bandstructure and NRL of the metawaveguide. CWL: continuous-wave laser; PC: polarization controller; BS: beam splitter; VOA: voltage-controlled attenuator; PD: photodetector. **b.** Scanning electron microscopy (SEM) image of the fabricated metawaveguide sample. **c.** Cross-section SEM image of the strip waveguide. **d.** Measured bandstructure and the corresponding interference patterns across the metawaveguide area at five different wavelengths. **e.** Measured and simulated forward and backward transmission spectra of a straight metawaveguide. **f.** Measured NRL of metawaveguide samples at various metawaveguide lengths.

To experimentally verify our theory, we fabricated the metawaveguide based on the designed parameters. The device was fabricated by depositing an a-Si thin film on a Ce:YIG film epitaxially grown on an sGGG substrate, followed by inductively coupled plasma-reactive ion etching (ICP-RIE) of the a-Si into the bowtie structure (Figs. 4b). The fabricated samples faithfully resemble our design in Fig. 2a (see Methods and supplementary information, section S3, for detailed discussion).

By illuminating two counter-propagating coherent continuous waves near a wavelength of 1550 nm into the metawaveguide (Fig. 4a) via single mode fibers, we observed interference pattern over the metawaveguide, from which we extracted the effective wavenumber and reconstructed the photonic bandstructure shown in Fig. 4d. The measured bandstructure shows a quasi-Dirac cone dispersion near 1525 nm (see measurement setup in supplementary information, section S6). Under zero applied magnetic field, the forward and backward transmission spectra of the metawaveguide overlap well (see supplementary information, section S6 and Fig. S29), confirming its reciprocity behavior and indicating negligible magnetic remanence in the Ce:YIG material. In contrast, under an applied magnetic field of 1000 Oe — sufficient to saturate the Ce:YIG — the backward and forward transmission spectra exhibited a peak and a valley near 1529 nm, respectively, corresponding to an NRL of 8.2 dB for a 400-μm-long metawaveguide (Fig. 4e). We further measured the NRL of metawaveguides of various lengths, from 50 to 400 μm (Fig. 4f). These measured NRLs increased linearly with waveguide length, yielding an NRL per unit length of 21 dB/mm at 1530 nm (see detailed propagation loss measurements in supplementary information, section S7). These results are consistent with our theoretical predictions (Fig. 2h and dashed lines in Fig. 4e).

We further characterized the NRPS of a 200-μm-long metawaveguide-inserted microring resonator (Figs. 5a and 5b). To measure the NRPS, we extracted the effective indices of the metawaveguide from the resonant frequencies of the resonator (see measurement setup in supplementary information, section S6). Based on the resonance wavelength shifts between forward and backward propagations, we calculated the NRPS of the metawaveguide (Fig. 5d), which exhibited a peak value of 47.78 rad/mm near 1579 nm — coinciding with the emergence of the zero-index state and topologically protected degeneracy point (inset of Fig. 5c; see detailed spectra in supplementary information, section S8). This result confirmed our theoretical prediction in Fig. 2f. This measured NRPS per unit length is 11.81 times greater than that of a state-of-the-art Si/Ce:YIG waveguide, which showed NRPS of 5.2 rad/mm

(see comparison in supplementary information, section S9). This giant NRPS under a weak magnetic field was enabled by the large momentum difference between the two branches (Fig. 2e). Additionally, we observed nearly overlapping resonance peaks for forward and backward transmission spectra near 1514 nm, outside of the zero-index wavelength, corresponding to the wavelength marked by star symbols in Fig. 2f. This result confirmed that the enhanced NRPS was localized near the zero-index wavelengths (see supplementary information, section S10, for further discussion).

We also characterized another figure of merit — defined as the NRPS divided by the propagation loss per unit length — considering both the NRPS and propagation loss of the metawaveguide. The propagation loss of the metawaveguide was calculated to be 15 dB/mm based on the measured quality factor of the microring. Therefore, our metawaveguide showed a figure of merit of 3.19 rad/dB, which exceeded that of a conventional a-Si/Ce:YIG channel waveguide (2 rad/dB[49]; see supplementary information, section S9, for discussion).

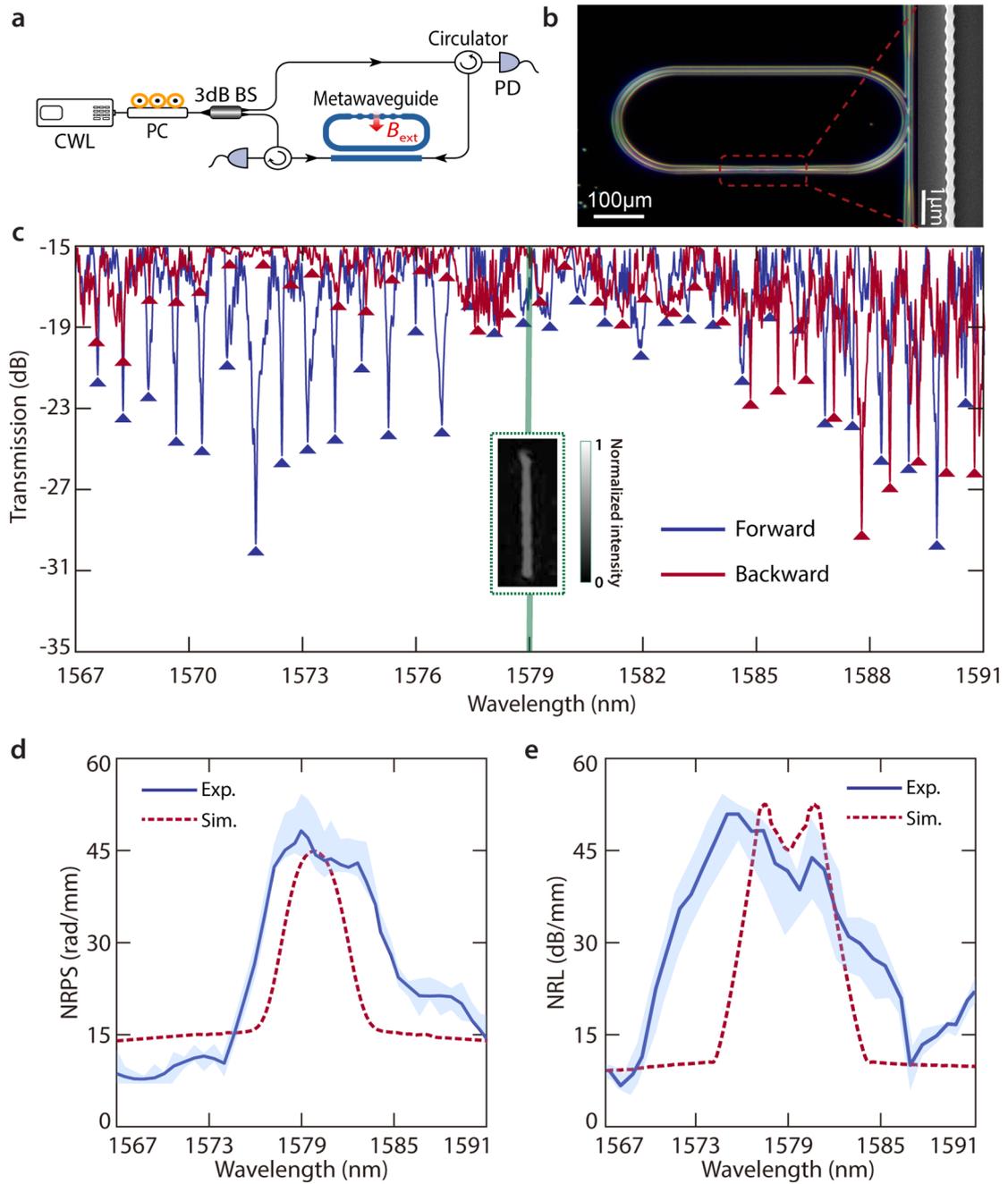

**Fig. 5. NRPS and NRL measurements of an a-Si/Ce:YIG metawaveguide-inserted microring resonator in the near infrared regime. a.** Schematic of the experimental setup for measuring NRPS and NRL in the microring resonator with a metawaveguide insert. CWL: continuous-wave laser; PC: polarization controller; BS: beam splitter. **b.** Optical micrograph and SEM image of the fabricated a-Si/Ce:YIG metawaveguide-inserted microring. **c.** Measured forward and backward transmission spectra of the metawaveguide-inserted microring. The inset shows a near-infrared image of the

interference pattern over the metawaveguide region at 1579 nm without an applied magnetic field, depicting an infinite spatial wavelength corresponding to the zero-index state. **d, e.** Measured and simulated NRPS (**d**) and NRL (**e**) spectra of the metawaveguide-inserted microring. The light blue shaded areas illustrate measurement uncertainties.

For the NRL, we observed significant differences in the amplitudes of the resonance peaks for forward and backward propagations, attributed to the nonreciprocal modulation of the intrinsic quality factor of the microring resonator caused by the NRL. Based on the extracted forward and backward intrinsic quality factors under a magnetic field of 1000 Oe, we calculated the NRL of the metawaveguide (Fig. 5e), showing a topological transition-induced peak value of 53.9 dB/mm near the zero-index wavelength of 1575 nm. This result is consistent with our theoretical prediction in Fig. 2h. Moreover, the measured NRL exceeded the highest values previously reported for conventional magneto-optical waveguides (see supplementary information, section S9, for discussion). It was also more than twice the NRL measured in the straight metawaveguide sample (Fig. 4f), which was attributed to fabrication-induced geometry difference between the straight metawaveguide sample and metawaveguide-inserted microring sample (see supplementary information, section S3, for discussion).

**Conclusion and discussion**

By transferring the concept of multi-value behavior-based arrow of time from the black hole horizon to electromagnetic systems, we designed a nonreciprocal non-Hermitian zero-index magneto-optical metawaveguide. By engineering the coupling between monopole and dipole modes, we introduced a ½-residue branch point into the complex eigenvalue space of the metawaveguide, generating a global multivalued feature. Even under an extremely weak magneto-optical response, our metawaveguide can still direct forward and backward electromagnetic waves to branches with largely distinct momenta and losses, leading to giant NRPS and NRL. We fabricated and tested a metawaveguide, exhibiting an NRPS of 47.78 rad/mm at 1579 nm — 11.81 times higher than that of state-of-the-art magneto-optical waveguides. Our metawaveguide also showed nonreciprocal "one-way" propagation behavior, featuring an NRL of 53.9 dB/mm at 1575 nm.

Although we demonstrated devices only in the near infrared regime, the principle of the multi-value behavior-based nonreciprocity is broadly applicable across the entire electromagnetic spectrum from microwave to visible, where waveguide-based magneto-optical devices are either bulky or non-existent. We simulated three metawaveguides operating at these frequencies (see supplementary information, section S11). Giant NRPS and NRL values were observed, demonstrating 2.3, 3.96, and 3.57 times shorter in effective length compared to the conventional magneto-optical devices at microwave, visible, and mid-infrared frequencies, respectively[55]. These metawaveguides showed shorter device lengths compared to their commercial bulk counterparts, despite part of the electromagnetic field being confined in dielectric waveguides without any magneto-optical response. These results demonstrated a promising solution to compact and integrated magneto-optical nonreciprocal waveguide devices across a wide electromagnetic spectrum.

In addition to magneto-optical nonreciprocity, our principle could also be transferred to nonreciprocity based on spatial-temporal modulation and nonlinear modulation, resulting in various compact nonreciprocal electromagnetic devices. Furthermore, our principle could be transferred to acoustic, elastic, and thermal systems, leading to compact isolators, circulators, and sensors for multiphysical fields. Moreover, we proposed a novel paradigm — geometry-based strong arrow of time in covariant and reversible physical systems, showing broad implications in many disciplines such as string theory, cosmology, and astronomy.

**Acknowledgements:**

**Funding:**

National Natural Science Foundation of China 62435009 (Y. L.)

National Natural Science Foundation of China 52450018 (L.B.)

National Natural Science Foundation of China U22A20148 (L.B.)

National Natural Science Foundation of China 52021001 (L.J.D.)

Sichuan Provincial Science and Technology Department 2025ZNSFSC0040 (L.B.)

Competitive Research Program Award from the NRF, Prime Minister's Office, Singapore NRF-CRP30-2023-0003 (C.-W.Q)

Project M22K2c0088 with A-8001322-00-00 from A*STAR MTC IRG, Singapore (C.-W.Q)

Beijing Municipal Natural Science Foundation Z220008 (Y.L.)

Ministry of Education, Republic of Singapore A-8002152-00-00 & A-8002458-00-00 (C.-W.Q)

Competitive Research Program Award from the National Research Foundation, Prime Minister's Office, Singapore NRF-CRP26-2021-0004 & NRF-CRP30-2023-0003 (C.-W.Q)

**Author contributions:**

Basic idea for the work: L.B., Y.L., Y.Y.L, Y.C.Y., and T.Y.Z.

Growing Ce:YIG film: Y.C.Y., T.C.Z., D.W., Z.X.W., D.D.F., and J.Q.

Theory: T.Y.Z.

Numerical simulations: Y.Y.L., Y.C.Y., and T.Y.Z.

Sample fabrication: Y.Y.L. and Y.C.Y.

Sample characterization: Y.Y.L.

Analyzation of the experimental results: Y.Y.L., Y.C.Y, and T.Y.Z.

Figures: Y.Y.L., Y.C.Y., T.Y.Z., T.D., Y.L., J.F.C, and P.L.

Supervising the research: L.J.D., C.-W.Q., L.B., and Y.L.

Manuscript: Y.Y.L., Y.C.Y., T.Y.Z., J.F.C, F.L.S., Y.L., L.B., and C.-W.Q.

**Competing interests:** Authors declare that they have no competing interests.

**Data and materials availability:** All data are available in the main text or the supplementary materials upon reasonable request from corresponding author.

**Supplementary Materials**

Materials and Methods

Supplementary information

Fig. S1 to S33

Tables S1 to S5

References (1–39)

Movies S1 to S4